\documentclass[12pt]{article}
\usepackage{amssymb}

\usepackage{amsmath}
\usepackage{bbold}
\usepackage{jheppub}

\def\bea{\begin{eqnarray}}
\def\eea{\end{eqnarray}}
\def\be{\begin{equation}}
\def\ee{\end{equation}}
\def\ms{\mathsf}
\def\td{\text{d}}

\def\bs{\boldsymbol}

\usepackage{braket}

\usepackage{tabularx}

\usepackage{booktabs}

\begin{document}
\begin{titlepage}
\setcounter{page}{1} \baselineskip=15.5pt \thispagestyle{empty}
\bigskip\
\begin{center}
{\Large \bf Testing Hamiltonian Reduced QED}
\vskip 5pt
\vskip 15pt
\end{center}
\vspace{0.5cm}
\begin{center}
 
{Thomas C. Bachlechner}

\vspace{1cm}

\textsl{Department of Physics, University of California San Diego, La Jolla, USA} 
\end{center}

{\small  \noindent  \\[1cm]
\begin{center}
	\textbf{Abstract}
\end{center}
Certain gauge transformations may act non-trivially on physical states in quantum electrodynamics (QED). This observation has sparked the yet unresolved question of how to characterize allowed boundary conditions for gauge theories. Faddeev and Jackiw proposed to impose Gauss' law on the action to find the Hamiltonian reduced theory of QED. The reduction eliminates the scalar gauge mode, renders the theory manifestly gauge invariant and the symplectic form non-singular. In this work we show that while the predictions of the reduced theory coincide with those of conventional QED for scattering events, it is experimentally distinguishable. Quantum interference of charges traveling along time-like Wilson loops that encircle (but remain clear of) electric fields is sensitive to a relative phase shift due to an interaction with the scalar potential. This is the archetypal electric Aharonov-Bohm effect and does not exist in the reduced theory. Despite its prediction over six decades ago, and in contrast to its well known magnetic counterpart, this  electric Aharonov-Bohm phenomenon has never been observed. We present a conclusive experimental test using  superconducting quantum interferometry. The Hamiltonian reduction renders a theta term non-topological. We comment on consequences for semi-classical gravity, where it may alleviate a problem with the measure.
\noindent}
\vspace{0.9cm}

\vfill
\begin{flushleft}
\small \today
\end{flushleft}
\end{titlepage}

\tableofcontents
\newpage
\section{Introduction}
The Maxwell action of electrodynamics, 
\be\label{maxwella}
S_\text{M}[A]=\int_{\cal M} -{1\over 2} \ms F\wedge * \ms F+A\wedge J\,,
\ee
with potential $A$, field strengths $\ms F\equiv dA$ and conserved charge current $J$ is not in general invariant under all gauge transformations $A\rightarrow A+\td \lambda$ at the boundary $\partial {\cal M}$. The dynamical equations, $d*\ms F=J$, are singular and contain pure gauge modes that necessitate an auxiliary gauge fixing condition. The gauge redundancy implies that the  action  principle is sufficient --- but not necessary --- to arrive at Maxwell's equations. Recently, much effort has focused on determining what (sometimes intricate) boundary conditions to impose on the potentials to render the variational problem well posed and the action finite \cite{Dirac:1955uv,Frohlich:1979uu,Balachandran:2013wsa,Strominger:2013lka,Balachandran:2014hra,Kapec:2015ena,Donnelly:2015hta,Hawking:2016msc,Strominger:2017zoo,Henneaux:2018gfi,Gomes:2018dxs,Gomes:2019xhu,Gomes:2019rgg,Gomes:2019xto,Harlow:2019yfa,Giddings:2019ofz}. Some progress has been made for specific configurations with antipodal matching of potentials and fields decaying sufficiently fast, and when the boundary is at null or spatial infinity. The issue remains controversial.

In this paper we are interested in  gauge invariant conditions at arbitrary boundaries, including non-vanishing  field strengths $F$ that are not usually considered. As suggested first by Schwinger \cite{Schwinger:1948yk,Schwinger:1959zz}, and made more explicit by Faddeev and Jackiw\footnote{This basic idea seems to have emerged independently in several distinct lines of work, for example in \cite{Khvedelidze:1994sd,Khvedelidze:2004my,Jezierski:1990vu,Chruscinski:1997ha}. See also \cite{Witten:1988hc} for an application to gravity.} \cite{Faddeev:1988qp,Jackiw:1993in},  we can impose Gauss' law at the level of the action to find the gauge invariant, Hamiltonian reduced action principle that implies both Maxwell equations without gauge fixing,
\be\label{maxwell}
\td*F=J\,,~~\td F=0\,.
\ee
Although the reduced theory contains no electric gauge potential, it is consistent with  observations. In the present work we will demonstrate that this theory is experimentally distinguishable from conventional quantum electrodynamics in a feasible quantum interference experiment probing time-like holonomy.

Fully gauge invariant boundary conditions are clearly impossible with the conventional Maxwell action (\ref{maxwella}), as its only degrees of freedom are contained in the  potential $A$ that transforms non-trivially under a shift by a total derivative. Instead, our starting point is the theory of a one-form $A$ and an independent two-form $F$ that are coupled to an electric current three-form $J$. This is the first-order action of electrodynamics, 
\be\label{actione}
S[A,F]=\int_{\cal M}  {1\over 2} F\wedge * F+{A\wedge(J-\td*F)}\,.
\ee
The variational problem is well posed for Dirichlet boundary conditions on the two-form $F$. In contrast to the Maxwell action $S_\text{M}[A]$, the first-order action $S[A,F]$ in (\ref{actione}) is invariant under all  transformations $A\rightarrow A+\td \lambda$ when the equations of motion are satisfied. Unfortunately, the symplectic form of (\ref{actione}) is singular and not yet canonical. Faddeev and Jackiw proposed a  procedure to eliminate the spurious degrees of freedom   by solving Gauss' law in (\ref{actione}) and performing a Darboux transformation to re-diagonalize the remaining symplectic form. The Gauss  law constraint reads $\bs \nabla\cdot\bs E=j^0$. Imposing such constraints can change the  equations of motion in a quantum theory of charged matter. For example, in Coulomb gauge a total derivative  $\bs \nabla\cdot(A_0\bs E)$ in the action would not contribute to the Euler-Lagrange equations, but imposing the  constraint  and thus replacing the total derivative term with $A_0j^0-\bs E^2$ would modify the Euler-Lagrange equation for matter when the boundary fields are non-zero.  Decomposing the two-form $F$ into general electric ($\bs E$) and magnetic ($\bs B= \bs \nabla\times \bs A_\text{T}$) fields and imposing Gauss' law we arrive at the action\footnote{This action first appeared  in \cite{Faddeev:1988qp,Jackiw:1993in} for vanishing boundary fields.}
\be\label{actionered}
S_\text{red}[\bs A_\text{T},\bs E_\text{T}]=\int_{\cal M} \td^4x ~\bs A_\text{T}\cdot \dot{\bs E}_\text{T}-\left[{\bs E^2(j^0)+\bs B^2(\bs A_\text{T})\over 2}-\bs A_\text{T}\cdot\bs j\right]\,.
\ee
The scalar potential has entirely disappeared from the action and we  refer to this theory as reduced electrodynamics. Bold  symbols denote spatial three-vectors, and $\bs A_\text{T}$ and $\bs E_\text{T}$ denote the  divergence-free transverse components of the vector potential and the electric field. The curl-free longitudinal mode   of the electric field is defined via Gauss' law and is an explicit function of the charge density $j^0$ and the boundary conditions. The action   now contains no gauge degrees of freedom and is fully invariant under $A\rightarrow A+\td \lambda$. No gauge fixing is necessary and the gauge invariant transverse components $(\bs A_\text{T}, {\bs E}_\text{T})$ form conjugate  phase space coordinates. 
In contrast to the conventional action (\ref{maxwella}), the reduced action  (\ref{actionered}) gives non-singular  dynamics that precisely agree with Maxwell's equations.

\begin{figure}
  \centering
  \includegraphics[width=.6\textwidth]{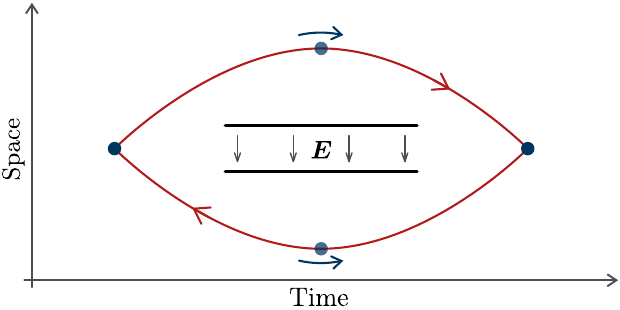}
  \caption{\small Illustration of the time-like Wilson loop (red) that is sensitive to the relative phase of a coherent particle beam traversing a field-free trajectory (blue arrows) around a capacitor containing time-dependent electric fields. This is the archetypal (type I) electric Aharonov-Bohm experiment.}\label{wilson}
\end{figure}
Although the equations of motion have changed, the reduction merely removed the gauge modes, so it is not obvious that reduced electrodynamics (\ref{actionered}) is physically distinguishable from conventional  electrodynamics (\ref{maxwella}). In this work we will demonstrate that the two theories are physically discernible in an experiment probing the time-like holonomy of quantum electrodynamics. It is well known that space-like and time-like Wilson loops are respectively sensitive to the transverse (magnetic) and scalar (electric)   potential in regions where the fields vanish, as in the Aharonov-Bohm experiments \cite{Ehrenberg_1949,PhysRev.115.485,PhysRev.123.1511}. While their better known, and experimentally verified \cite{1960PhRvL...5....3C,1962NW.....49...81M}, magnetic effect is sensitive only to the gauge invariant transverse potential $\bs A_\text{T}$, their electric effect is sensitive to the gauge covariant scalar potential.   This electric  memory \cite{Bieri:2013hqa,Susskind:2015hpa,Pasterski2017,Hamada:2017bgi} effect describes the relative phase shift of particles traversing a Wilson loop  that winds around, but never enters a non-vanishing time dependent electric field, as illustrated in Figure \ref{wilson}. 
   The reduced theory  contains no scalar gauge potential, and correspondingly we will find that it would be inconsistent with the electric Aharonov-Bohm phenomenon. Perhaps surprisingly, this archetypal electric Aharonov-Bohm effect has never been observed.\footnote{Some experimental studies tested and failed to verify the electric AB effect \cite{PhysRevB.40.3491}, while others made positive observations  \cite{mysteryreference,cite-key,PhysRevB.67.033307} in configurations referred to as ``type II'' where --- in contrast to the in the archetypal ``type I'' AB experiments --- the charges traverse the non-vanishing electric field due to the bias voltage. We will demonstrate in detail below that these observations are indeed consistent with the reduced theory.} In \cite{Bachlechner:2019deb} we propose a conclusive experiment using quantum interference of superconducting order parameters.

This paper is organized as follows. We begin in \S\ref{firstorder} with a review of the first-order action for electrodynamics. In \S\ref{jf} we discuss the Hamiltonian dynamics of constrained systems, following the phase space reduction advocated by Faddeev and Jackiw. The phase space  of the action (\ref{actione}) is not manifest, so   we carefully decompose the fields in order to exhibit the symplectic structure and to discriminate between gauge and physical degrees of freedom. We show that on-shell the first order action (\ref{actione}) coincides with the action for reduced electrodynamics (\ref{actionered}) \cite{Faddeev:1988qp,Jackiw:1993in}. In order to allow a comparison with the standard model, we couple the reduced action  to a Dirac fermion in \S\ref{sqed}. For vanishing boundary field strengths we will recover the  conventional  Lagrangian of spinor quantum electrodynamics. To study an experimental test of the reduced theory, in \S\ref{sqed} we discuss a charged non-relativistic scalar  that  describes Cooper pair condensates within superconductors. We derive a modified Ginzburg-Landau equation in \S\ref{gltheory} that predicts a magnetic, but no electric Aharonov-Bohm effect. The Josephson effects remain operational. In contrast to QED, the path integral measure is trivial in the canonical reduced theory. This observation allows to immediately approximate the semi-classical path integral by the exponent of the reduced on-shell action, as we discuss in \S\ref{semiclassicalr} at the example of electron pair production. We briefly comment on consequences for the semi-classical path integral in gravity, where the Hamiltonian reduction alleviates a problem with the measure. We conclude in \S\ref{conclusions}.

\section{The first-order action for electrodynamics}\label{firstorder}
To prepare our discussion of the Hamiltonian reduction in the next section, let us review some properties of the first-order action (\ref{actione}) for conventional electrodynamics.

\subsection{Boundary conditions and symplectic form}
The first-order action $ S [A,F]$ in (\ref{actione}) is a functional of a  two-form $F$ and an independent one-form $A$, given some closed current three-form $J$. The variation of the action is
\be\label{actione2}
\delta S [A,F] =\int_{\cal M} *\delta F\wedge(F-\td A)+\delta A(J-\td*F)-\int_{\partial \cal M}A\wedge*\delta F\,.
\ee
The Euler-Lagrange equations follow from the variational principle $\delta S=0$ under gauge invariant boundary conditions on the field strengths $F$. Specifically, the boundary term in the variation of the action vanishes when the electric field perpendicular\footnote{This includes the electric fields on the Cauchy surface.}, and the magnetic field parallel to the boundary $\partial \cal M$ are fixed. These boundary conditions fix the external electric fields and the electric charge contained within $\cal M$. For these gauge invariant boundary conditions, the variational principle yields the inhomogeneous Maxwell equations and the Bianchi identity
\be\label{eomelectric}
J=\td*F\,,~~\td A=F\,.
\ee
The   equations (\ref{eomelectric}) are sufficient but not necessary to arrive at the homogeneous Maxwell equation $\td F=d^2A=0$, as they contain a zero mode in the form of the gauge transformation $A\rightarrow A+\td \lambda$. We will see in the next section that imposing Gauss' law on the first-order action eliminates this zero mode from the phase space.

The symplectic two-form on phase space is given by
\be
\Omega=\int_\Sigma \delta A\wedge *\delta F\,,
\ee
where $\Sigma$ is a purely spatial three-dimensional slice of $\cal M$.  On the surface $\td A=F$, the symplectic form coincides with that of the  Maxwell theory (\ref{maxwella}), and contains zero modes associated with  gauge transformations. Since the symplectic form is singular, it is not possible to solve the dynamical equations (\ref{eomelectric}). Conventionally, the dynamical equations are supplemented with an auxiliary gauge fixing condition that allows to solve the boundary value problem. 

\subsection{Expanding the first order action}
In order to move towards a Hamiltonian formulation for electrodynamics, we now expand the two-form $F$ and the one-form $A$ in terms of three-vectors. In particular, we define the components
\be
F^{\mu\nu}=\left(
\begin{array}{cccc}
 0 & -E^1 & -E^2 & -E^3 \\
 E^1 & 0 & -B^3 & B^2 \\
 E^2 & B^3 & 0 & -B^1 \\
 E^3 & -B^2 & B^1 & 0 \\
\end{array}
\right)^{\mu\nu}\,,~~A^\mu=\left(
\begin{array}{c}
 A^0\\
 \bs A
\end{array}
\right)^{\mu}\,,
\ee
where bold symbols denote upper-index three-vectors, for example $(\bs A)^i\equiv A^i=-A_i$. In terms of these components we have the first-order action
\bea
S &=&\int_{\cal M} \td^4x~  -A_\mu (j^\mu+\partial_\nu {\ms F}^{\mu\nu})+{1\over 4} {\ms F}_{\mu\nu}{\ms F}^{\mu\nu}\,\nonumber\\
&=&\int_{\cal M} \td^4x~-A_0(j^0-\bs\nabla\cdot \bs E)+\bs A\cdot(\bs j-\bs\nabla\times \bs B+\dot{\bs E})+{1\over 2}({\bs B^2-\bs E^2})\label{expandedaction}\,.
\eea
The kinetic term of this theory is $\bs A\cdot \dot{\bs E}$, i.e. the three-potential $\bs A$ and the electric field $\bs E$ form conjugate pairs of phase space coordinates. These coordinates  include a longitudinal, pure gradient gauge mode of $\bs A$ that will disappear when we impose Gauss' law $j^0=\bs \nabla\cdot \bs E$ on the action.

\subsection{Comparison to the  Maxwell action}
We can compare to the conventional Maxwell action by restricting the first-order action onto the surface $F=\td A$, where we find the relative boundary term
\be
S[A,F]\big|_{F=\td A}=S_\text{M}[A]+\int_{\cal M} \td(A\wedge *\td A)\,.
\ee
The boundary term renders the on-shell action gauge invariant and ensures that the variational problem is well posed for gauge invariant boundary conditions on $F$, see also \cite{Brown:1987dd,Brown:1988kg,Brown:1997dm,Hawking:1995ap}. In general the boundary term is non-vanishing for classical configurations, but it vanishes when the boundary fields are zero (or decay sufficiently fast).

We can intuitively understand the appearance of the non-trivial boundary term by recalling that the Maxwell action has a well posed variational problem for boundary conditions that fix the perpendicular magnetic and parallel electric fields, i.e. for fixed magnetic charges. Correspondingly, it is easy to check that the Maxwell action on-shell coincides with the magnetically dual first-order action. We are primarily interested in describing electric charges, so the Maxwell action does not obviously apply. This has important and well-known consequences for semi-classical physics that we discuss in \S\ref{semiclassicalr}.

\section{Hamiltonian  reduction}\label{jf}
In this section we review the method of Hamiltonian reduction that  Faddeev and Jackiw suggest to determine the physical phase space of constrained systems. We closely follow \cite{Jackiw:1993in}. The  basic idea is to impose all true constraints (such as Gauss' law) and perform Darboux transformations on the configuration space to eliminate the remaining spurious degrees of freedom from the action. For electromagnetism this results in an  action that yields the non-singular Maxwell equations of motion (\ref{maxwell}). In contrast,  Dirac's conventional approach  introduces  gauge fixing conditions to evaluate the (otherwise singular) dynamical equations (\ref{eomelectric}),  including solutions for Lagrange multipliers.

We will explicitly apply Hamiltonian reduction  to the example of  electromagnetism with classical sources to see how the longitudinal gauge mode is eliminated. Although the dynamical equations manifestly differ between the two methods to construct a phase space, it is not obvious that this change has observational consequences. We defer a discussion of observable differences to \S\ref{gltheory}, where we will see that the elimination of the scalar potential affects the  electric Aharonov-Bohm effect as this phenomenon  is sensitive to the existence of a scalar potential.

\subsection{Dirac approach to constrained systems}
Dirac's  prescription for the construction of a phase space  can be summarized as follows: consider a Lagrangian $L(q^i,\dot{q}^i)$ describing $N$ degrees of freedom $q^i$, labeled by $i=1\dots N$, perform a Legendre transform that yields the Hamiltonian in terms of momenta $p^i$, $H(p^i,q^i)=p_i\dot{q}^i-L$, postulate canonical brackets between the pairs $(p^i,q^i)$, and study the dynamics generated by the Hamiltonian equations. This prescription is straightforward when the dynamics are non-singular and it is possible to solve for the velocities in terms of the canonical coordinates. 

To make the prescription more explicit we introduce the $2N$ phase space coordinates $\zeta^I$, labeled by $I=1,\dots,2N$ containing the momenta $p_i$ and generalized coordinates $q^i$, labeled by $i=1,\dots N$, to write the Lagrangian as
\be
L={1\over 2}\zeta^I\omega_{IJ} \dot{\zeta}^J-H(\zeta)\,.
\ee
We dropped a total time derivative that does not affect the dynamics and the matrix $\omega$ is anti-symmetric. The first term in the Lagrangian one-form $L \td t$ defines the canonical one-form potential, $a\equiv {1\over 2}\zeta^I\omega_{IJ} \delta\zeta^J$, with an associated symplectic two-form $f=\delta  a={1\over 2}\delta\zeta^I\omega_{IJ} \delta\zeta^J$. The variational principle $0=\delta \int \td t L$ yields the dynamical equations
\be
\omega_{IJ}\dot{\zeta}^J={\partial H\over \partial \zeta^I}\,.
\ee
Whenever the symplectic form is non-singular the inverse $\omega^{IJ}$ exists, so we can solve for the velocities
\be
\dot{\zeta}^J=\omega^{IJ}{\partial H\over \partial \zeta^I}\,,
\ee
and define the brackets $\{\zeta^J,\zeta^I\}=\omega^{IJ}$. For the simple unconstrained theory above, this gives the standard relations $\{q^i,p^j\}=\delta^{ij}$.

The presence of spurious variables in the action, such as Lagrange multipliers, can render the symplectic two-form singular. In this case there are not sufficient dynamical equations to solve the momenta for all $N$ velocities. In Dirac's language this means that there exist first-class constraints whose  brackets with all other constraints vanish. In order to proceed, the conventional approach then instructs to supplement the theory with gauge fixing conditions that allow a solution for all coordinates via the supplemented dynamical equations. The new system is non-singular and can provide a solution that includes the (previously) spurious variables, such as Lagrange multipliers.

Faddeev and Jackiw propose an alternative method that yields identical commutation relations as Dirac's procedure, but rather than supplementing arbitrary dynamical equations for the spurious degrees of freedom, the unphysical\footnote{To be precise, we call a degree of freedom ``unphysical'' whenever its dynamics are not provided by the variational principle. For example,  Lagrange multipliers are unphysical in this sense. Whether or not these degrees of freedom exist in nature is an open question that can be answered by experiments, see \S\ref{gltheory}.} variables are eliminated, as we now review.

\subsection{Faddeev-Jackiw approach to constrained systems}\label{FJacs}
Consider a general first-order Lagrangian of the form
\be
L=a_I(\zeta)\dot{\zeta}^I-H(\zeta)\,,
\ee
where the one-form potential $a=a_I(\zeta) \delta \zeta^I$ depends arbitrarily  on the $K$ variables $\zeta^I$. The Euler-Lagrange equations
\be
f_{IJ}(\zeta) \dot{\zeta^J}={\partial\over \partial \zeta^J} H(\zeta)\,,
\ee
contain only the non-singular components of the symplectic two-form $f=\delta a={1\over 2} f_{IJ}(\zeta)\delta\zeta^I \delta\zeta^J$, and therefore may not allow for a solution of all variables $\zeta^I$. We begin by performing a coordinate transformation $\zeta\rightarrow \tilde{\zeta}$, that projects onto the $2N=K-N'$ dimensional non-singular subspace of $f$, and denote by $z$ the remaining $N'$ zero modes of $f_{IJ}$. Dropping the tilde for compactness gives a Lagrangian of the form
\be
L=a_I(\zeta)\dot{\zeta}^I-H(\zeta,z)\,,
\ee
where the new two-form $f=\delta a$ is now non-singular. The existence of a non-singular symplectic two-form $f$ allows us to apply Darboux's theorem which guarantees the existence of a coordinate transformation $Q(\zeta)$, such that the one-form potential reads
\be
a={1\over 2} Q^k\omega_{kl}{\delta Q^l}\,,~~\bs \omega=\left(\begin{array}{cc}0&\bs 1\\- \bs 1&0 \end{array}\right)\label{omega}\,,
\ee
where $\bs 1$ is the $N\times N$ unit matrix and $Q^k$ are the canonical coordinates. The  variables $z$ that do not appear in the symplectic form are non-dynamical, but may be physical. These $N'$ variables $z^l$ are not part of the physical phase space and subject to the $N'$  constraint equations
\be
{\partial H(\zeta,z)\over \partial z_l}=0\,,~~~~l=1,\dots,N'\,.\label{hamconstrs}
\ee
For variables appearing non-linearly in $H$, these constraints can yield solutions $z^l(\zeta)$ that we  substitute into the Lagrangian without affecting the symplectic form. For variables appearing linearly in $H$, however, there exist no solutions. These are the unphysical variables of the system that generate the only true constraints  of the form $\xi(\zeta)=0$. Imposing the associated constraints on the Lagrangian eliminates the unphysical variables and potentially changes the symplectic form and equations of motion, as it establishes relations between the dynamical variables $\zeta$. These new relations can reduce the number dynamical of degrees of freedom below $2N$. This last step is crucial, since a change of the symplectic form can affect the dynamics of a system. The  prescription is iterated until one arrives at an unconstrained Lagrangian containing only physical degrees of freedom. Out of the $K$ original degrees of freedom this procedure gives solutions for all non-dynamical, but physical variables and eliminates all unphysical variables from the theory. This is in stark contrast to Dirac's procedure, where the supplemental constraints give solutions for all variables, including the unphysical ones.

The functional integral for the reduced quantum theory is given by
\be
Z=\int \Pi_i {\cal D}\zeta^i \sqrt{|f_{kl}|} e^{i\int \td t L(\zeta)}\,,
\ee
which contains the trivial measure when the integration is performed over the  coordinates for which the symplectic form is canonical, as in (\ref{omega}).

The Hamiltonian reduction  changes the dynamical equations relative to Dirac's prescription. In the following sections we discuss the consequences for quantum electrodynamics. We will show that perhaps surprisingly, the observable consequences are consistent with current experiments and will be tested in the near future.

\subsection{Application to classical electrodynamics}
To illustrate the consequences of the Hamiltonian reduction we now consider its application to the classical theory (\ref{actione}) describing electrodynamics with charges and fixed boundary field strengths. Since the symplectic form contains an integral over the constant time slices $\Sigma$, we begin with the action (\ref{expandedaction}), where we expanded the one- and two-forms into temporal and spatial components,
\be
S [\bs B, \bs E,A_\mu]=\int_{\cal M} \td^4x~-A_0(j^0-\bs \nabla\cdot \bs E)+\bs A\cdot(\bs j-\bs \nabla\times \bs B+\dot{\bs E})+{1\over 2}({\bs B^2-\bs E^2})\label{electricaction}\,.
\ee
This action initially contains $10$ variables: a Lagrange multiplier $A_0$, as well as two transverse and one longitudinal mode for each of the  spatial three-vectors $\bs A$, $\bs E$ and $\bs B$. For now we are interested in the classical theory, so the conserved current four-vector $j^\mu$ describes a prescribed electric source and we defer the discussion of dynamical matter to \S\ref{sqed} and \S\ref{gltheory}. Both the scalar potential $A_0$ and the magnetic field $\bs B$ appear in terms without time derivatives and therefore are not themselves dynamical. To be concrete we take the spacetime region ${\cal M}$ as the product of a time interval ${\cal I}$ with a compact subset $\Sigma$ of three-dimensional Euclidean space, ${\cal M}=I\times \Sigma$. On the spatial manifold we can then perform a Hodge decomposition of the spatial vector components into longitudinal and transverse   modes,
\be
\bs A\equiv \bs A_\text{T}+\bs \nabla A_\text{s}\,,~~\bs E\equiv \bs E_\text{T}+\bs \nabla E_\text{s}\,,
\ee
where $\bs \nabla \cdot \bs A_\text{T}=\bs \nabla \cdot \bs E_\text{T}=0$, and $A_\text{s}$ and $E_\text{s}$ denote scalar potentials. Integration by parts\footnote{We drop the total derivative term $\int_{\Sigma} \td^3x (E_\text{s}\bs A_\text{T} +\bs A_\text{T}\times \bs B)$ that does not affect the dynamics.} and noting that the boundary conditions  satisfy the equations of motion yields the Lagrangian 
\be
L  [\bs B, \bs E,A_\mu]=\int_{\Sigma} \td^3x~{1\over 2}(\bs A_\text{T}\cdot \dot{\bs E}_\text{T}-\dot{\bs A}_\text{T}\cdot {\bs E}_\text{T})-H(\bs A_\text{T},A_\text{s},A_0,\bs E_\text{T},E_\text{s},\bs B)\,,
\ee
where we extracted a canonical  kinetic term. The Hamiltonian reads
\be
H=\int_{\Sigma}\td^3x{\bs E^2-\bs B^2+2 \bs \nabla \times \bs A_\text{T}\cdot\bs B \over 2}+A_0(j^0-\bs \nabla\cdot \bs E)+A_\text{s}(\bs \nabla \cdot \bs j+\bs \nabla \cdot \dot{\bs E})-\bs A_\text{T}\cdot\bs j\,,
\ee
but it still contains time-derivatives. Following the Hamiltonian reduction, we can solve (\ref{hamconstrs}) for the magnetic field in terms of the dynamical variable $A_\text{T}$,
\be
\bs B(\bs A_\text{T})=\bs \nabla \times \bs A_\text{T}\,.
\ee
Solving for the magnetic field does not change kinetic terms or the symplectic form. On the other hand, $A_0$ appears linearly in the Lagrangian and is absent from the dynamical equations. This unphysical variable has to be eliminated. The variation with respect to $A_0$ yields Gauss' law which has the solution
\be\label{essol}
E_\text{s}(j^0)=-{1\over 4\pi}\int_{\Sigma}\td^3x' {j^0\over |\bs x-\bs x'|}+{1\over 4\pi}\int_{\partial \Sigma}\td \bs S'\cdot {\bs E\over |\bs x-\bs x'|}\,,
\ee
where the second term is determined by the boundary conditions that fix the electric field perpendicular to the boundary. Substituting (\ref{essol}) and   the continuity equation\footnote{This step is not required in the quantum theory, where the longitudinal gauge mode can be absorbed by a Darboux transformation of the matter fields.} $\partial_\mu j^\mu=\bs \nabla \cdot \bs j+\bs \nabla \cdot \dot{\bs E}=0$ in the Hamiltonian, we see that the longitudinal gauge mode $A_\text{s}$ disappears, and we finally arrive at the Hamiltonian of the reduced theory
\be
H_{\text{red}}(\bs E_\text{T},\bs A_\text{T})=\int_{\Sigma}\td^3x{\bs E^2(j^0)+(\bs \nabla \times \bs A_\text{T})^2\over 2}-\bs A_\text{T}\cdot\bs j\,.\label{redham}
\ee
The symplectic form is now canonical, with the $4$ dynamical variables consisting of the transverse modes of both the electric field and the vector potential. While the magnetic field $\bs B$ and the  longitudinal mode of the electric field  $\bs E$ are non-dynamical, these fields are physical since their evolution is provided by the equations of motion. In contrast, the scalar modes $A_s$ and $A_0$ are neither dynamical nor physical. We immediately recognize the  canonical Hamiltonian (\ref{redham}) as the total energy of the system within the compact spatial region $\Sigma$. This is famously not the case for the Hamiltonian of the conventional Maxwell action (\ref{maxwella}), which in general does not agree with the energy.\footnote{It is easy to verify that the Hamiltonian (\ref{redham}) differs from the Hamiltonian of the conventional theory by a boundary contribution that is in general non-vanishing.}

To summarize, out of the original 10 degrees of freedom, we arrive at a four dimensional dynamical phase space (containing the two propagating photon polarizations and their momenta), four physical but non-dynamical variables (the longitudinal electric field and the magnetic field), and two non-dynamical and unphysical variables (the longitudinal vector potential and $A_0$). There are less physical variables than in the conventional Maxwell theory, where solutions for $A_0$ and $\bs \nabla A_\text{s}$ exist. It is perhaps entertaining to note that reduced QED is a rare theory of physics beyond the standard model in which we do not add towards, but instead reduce the number of fields.

The only remaining dynamical degrees of freedom are the gauge invariant transverse modes $ \bs A_\text{T}$ and $ \bs E_\text{T}$ with reduced Lagrangian
\be\label{Lred}
L_\text{red} [\bs E_\text{T},\bs A_\text{T}]=\int_{\Sigma} \td^3x~{1\over 2}\bs \zeta^\top\bs \omega\dot{\bs \zeta}-H_\text{red}(\bs E_\text{T},\bs A_\text{T})\,,
\ee
where $\zeta^\top=(\bs A_\text{T}^\top,\bs E_\text{T}^\top)$, and the symplectic matrix $\bs \omega$ is canonical, as in (\ref{omega}). The variational principle $0=\delta \int \td t ~L_\text{red}$ yields all Maxwell equations (but not all components of the Bianchi identity).

The idea that only the transverse, gauge invariant potentials are physical has existed long before Faddeev and Jackiw. For example, Schwinger emphasized that the Coulomb (or ``radiation'') gauge is superior to manifestly covariant gauges, and that the  scalar gauge modes can be eliminated from the equations of motion by solving Gauss' law   \cite{Schwinger:1948yk}. In \cite{Schwinger:1959zz} Schwinger proposed the gauge invariant canonical one-form of the reduced system (\ref{Lred}), but they did not discuss the observable effects.

\section{Spinor electrodynamics}\label{sqed}
In this section we will discuss an electrically charged Dirac spinor $\psi$  in reduced quantum electrodynamics and show that the predictions for  scattering experiments agree with those of  QED.

\subsection{Lorentz invariance}\label{lorentzinv}
Let us briefly recall the deep connection between Lorentz and gauge invariance for charged matter \cite{Weinberg:1995mt}. Weinberg demonstrated that it is not possible to construct a Lorentz four-vector $A_\mu$ from creation and annihilation operators for massless spin 1 particles \cite{1964PhL.....9..357W}. Instead, the four-potential $A_\mu$ transforms under Lorentz transformations $\Lambda$ as a four-vector only up to a total derivative, 
\be\label{ashift}
A_\mu\rightarrow \Lambda_\mu^{~\nu} A_\nu+\partial_\mu\lambda\,.
\ee
To retain Lorentz invariance of the interactions, we then require that the action for couplings to matter $S_\text{matter}$ be invariant under (\ref{ashift}),
\be\label{lorentzreq}
0={\delta S_\text{matter}}=\int_{\cal M}\td^4x {\delta S_\text{matter}\over \delta A_\mu}\partial_\mu \lambda\,.
\ee
For theories of matter $\psi$  that are invariant under a global transformation, $\psi\rightarrow e^{-i\alpha}\psi$, the matter action  transforms under local\footnote{By ``local'' in this work we mean to include transformations that do not vanish at boundaries.} transformations $\alpha=\alpha(x^\mu)$ as
\be
{\delta S_\text{matter}}={\hbar\over q}\int_{\cal M}\td^4x j^\mu\partial_\mu \alpha(x^\mu)\,.
\ee
The simultaneous infinitesimal transformations $\delta \psi=i\alpha q/\hbar\psi$ and $\delta A_\mu=\partial_\mu \lambda$ then give the combined variation
\be
{\delta S_\text{matter}}=\int_{\cal M}\td^4x {\delta S_\text{matter}\over \delta A_\mu}\partial_\mu \lambda+{\hbar\over q}j^\mu\partial_\mu \alpha(x^\mu)\,.
\ee
There are now two ways to satisfy or restore the requirement (\ref{lorentzreq}) for Lorentz invariance: (A) either the variation of the matter action with respect to a gauge transformation vanishes and matter transforms under a global symmetry $\alpha(x)=\alpha$, or (B) the potential couples to the conserved current via $A_\mu j^\mu$ and matter transforms under the joint local transformation $\alpha(x)=\lambda(x)$,
\be
\text{(A)}:~\int_{\cal M}\td^4x {\delta S_\text{matter}\over \delta A_\mu}\partial_\mu \lambda=0\,,~~\alpha(x)=\alpha\,,~~\text{(B)}:~{\delta S_\text{matter}\over \delta A_\mu}=-j^\mu\,,~~\alpha(x)={ q\over \hbar}\lambda(x)\,.
\ee
Either choice will result in  Lorentz invariant interactions.\footnote{This does not preclude spontaneous breaking of Lorentz invariance.} The former option retains the global symmetry of $\psi$, while the latter option promotes the global to a local symmetry and introduces Planck's constant in the transformation law of a spacetime symmetry. 
In  QED option (B) is realized in order to maintain Lorentz invariance. We can already anticipate that since reduced electrodynamics is manifestly gauge invariant, it realizes option (A), which also ensures Lorentz invariant interactions.

\subsection{Hamiltonian reduction of spinor electrodynamics}

The Dirac spinor $\psi$ contains no constrained degrees of freedom, so we do not have to repeat the  entire Hamiltonian reduction to find the action. Instead, we can add the Dirac Lagrangian $L_{\frac{1}{2}}=\int_{\Sigma}\td^3x \bar{\psi}(i\gamma^\mu\partial_\mu-m)\psi$ to the reduced Lagrangian for electrodynamics (\ref{Lred}). Substituting the conserved current $j^\mu=e \bar{\psi}\gamma^\mu \psi$ we then immediately find the reduced action
\be\label{electricd}
L_{\frac{1}{2}}=\int_{\Sigma} \td^3x~\bs A_\text{T}\cdot \dot{\bs E}_\text{T}+i{\psi}^\dagger\dot{\psi}-H_\text{red}(\bs E_\text{T},\bs A_\text{T},i\psi^\dagger,\psi)\,,
\ee
where
\be\label{hamred}
H_{\text{red}}=\int_{\Sigma}\td^3x{\bs E^2(\psi^\dagger\psi)+(\bs \nabla \times \bs A_\text{T})^2\over 2}-\bar{\psi}\left(\bs \gamma\cdot \left[i\bs \nabla+ e \bs A_\text{T}\right]-m\right)\psi\,.
\ee
We perform the full reduction of spinor quantum electrodynamics in Appendix \ref{app1} to explicitly confirm the shortcut we used to arrive at (\ref{electricd}).

We can further manipulate the Hamiltonian (\ref{hamred}) for the special case of scattering of charged particles. For scattering events of massive, localized particles in asymptotically flat space we can choose the spatial boundary such that the particles stay far within the spatial region $\Sigma$ and we can pick an integration constant in (\ref{essol}) such that the electric potential vanishes as $E_\text{s}\sim \mathcal{O}(1/|\bs x|)$. Integration by parts gives
\be\label{hamredzerofield}
H_{\text{red}}=\int_{\Sigma}\td^3x{\bs E^2_\text{T}+(\bs \nabla \times \bs A_\text{T})^2\over 2}-{E_\text{s} j^0\over 2}-\bs J\cdot \bs A_\text{T}-\bar{\psi}\left(i\bs \gamma\cdot \bs \nabla-m\right)\psi+\int_{\partial \Sigma}\td \bs S\cdot ( E_\text{s} \bs \nabla E_\text{s})\,.
\ee
Employing Dirac brackets with the definition $\bs \nabla\cdot \bs A_\text{T}\equiv 0$, we find the standard commutation relations for the transverse modes,
\bea
&&\left[E_{\text{T},i}(\bs x),A_\text{T}^j(\bs y)\right]=i\delta_{\,i}^j \delta^3(\bs x-\bs y)+i{\partial^2\over \partial x^j\partial x^i}\left({1\over 4\pi |\bs x-\bs y|}\right)\,,\nonumber\\&&\left[E_{\text{T},i}(\bs x),E_{\text{T},j}(\bs y)\right]=\left[A_\text{T}^i(\bs x),A_\text{T}^j(\bs y)\right]=0\,.
\eea
For localized charged particles in asymptotically flat space the integrand in the boundary term in the Hamiltonian vanishes as $\propto \mathcal{O}(1/|\bs x|^3)$, so the boundary integral vanishes. Under these conditions, we recognize that the the reduced Hamiltonian (\ref{hamredzerofield}) coincides with the conventional  Hamiltonian for QED in Coulomb gauge \cite{Weinberg:1995mt}, and so the  predictions will agree when the boundary fields vanish.

\subsection{Gauge invariance}
We already anticipated in \S\ref{lorentzinv} that reduced quantum electrodynamics will retain the original invariance under the global $U(1)$ symmetry of the Dirac spinor, $\psi\rightarrow e^{i\alpha}\psi$, as well as the gauge invariance of classical electrodynamics, $A\rightarrow A+\td \lambda$. While the theory is not invariant under arbitrary time dependent local $U(1)$ transformations at the boundary, examination of the theory (\ref{electricd}) reveals  invariance under all {\it spatial} local $U(1)$ transformations, i.e.
\be\label{spatialu1}
\{\psi,\,\bs A_\text{T}\}\rightarrow \{e^{i e \lambda(\bs x)}\psi,\,\bs A_\text{T}+\bs \nabla \lambda(\bs x)\}\,.
\ee
We can therefore define a gauge invariant derivative that is covariant under a re-definition of the field-decomposition, $\bs D\equiv \bs \nabla-ie \bs A_\text{T}$. In reduced quantum electrodynamics there is no need to promote the global symmetry of matter to a local symmetry. This means that --- in contrast to QED --- Planck's constant does not appear in a  spacetime symmetry transformation.

\section{Testing reduced quantum electrodynamics with  Wilson loops}\label{gltheory}

We have seen in the previous section that reduced quantum electrodynamics reproduces the conventional theory when the boundary electric fields vanish. We now discuss the case of non-vanishing boundary fields, that will allow to experimentally distinguish between the conventional and the Hamiltonian reduced theories of quantum electrodynamics. In the reduced theory we eliminated the scalar mode of the gauge potential from the dynamical equations, so we are interested in testing the significance of this electric potential in regions of vanishing field strengths.

Aharonov and Bohm proposed two experiments that were designed to directly verify the physical significance of the magnetic and electric potentials in the quantum theory, as we will review in \S\ref{abe}. While the magnetic effect that is mediated by the transverse mode of the vector potential has been observed, the corresponding electric effect that is sensitive to the longitudinal mode of the vector potential has never been verified.\footnote{In this paper we exclusively refer to the field independent Aharonov-Bohm effect, sometimes referred to as ``type I''  \cite{1996fqml.conf....8A}. In contrast, a field dependent electric AB effect  has been observed \cite{cite-key,PhysRevB.67.033307}. This phenomenon, where the charges traverse non-vanishing electric fields, is sometimes referred to as ``type II'', and does not suffice to verify the independent physical significance of the potential. We will see below that reduced QED predicts a type II, but no type I electric AB effect.} The original experimental proposal involved the interference of electron beams, that turns out to be challenging to realize. Instead, we will employ the simpler setup of two superconductors that interact via a Josephson junction. To study this system, we begin in \S\ref{nrcs} with a  discussion of a non-relativistic charged scalar in reduced quantum electrodynamics and identify a gauge invariant voltage. In \S\ref{supconds} we identify the charged scalar with the Cooper pair condensate of a superconductor, review basic superconductor physics and derive some relevant results from the reduced theory, including the London equation and  Josephson relations for superconductor junctions. Finally, in \S\ref{ABJ}, we discuss a new superconductor interference experiment \cite{Bachlechner:2019deb} that is sensitive to the physical significance of the scalar gauge potential.

\begin{figure}
  \centering
  \includegraphics[width=1\textwidth]{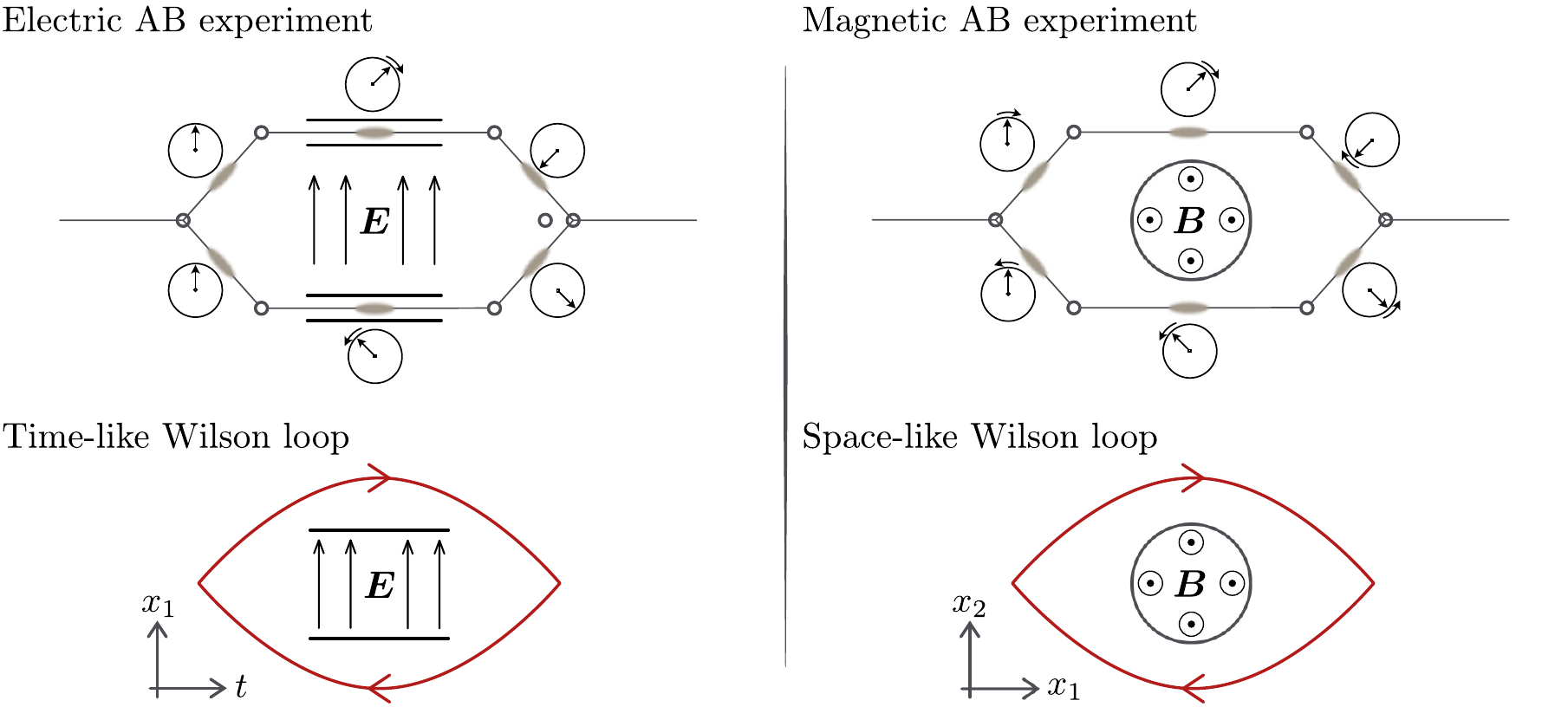}
  \caption{\small Schematic illustration of both the electric (left) and magnetic (right) Aharonov-Bohm (AB) experiments. The observable relative phase of two particles depends on a closed time-like Wilson loop, and a closed space-like Wilson loop, respectively.}\label{ABeffects}
\end{figure}

\subsection{The Aharonov-Bohm experiments}\label{abe}

Aharonov and Bohm (AB) suggested two quantum interference experiments aimed at verifying the physical significance of the longitudinal and transverse modes of the gauge potential \cite{Ehrenberg_1949,PhysRev.115.485,PhysRev.123.1511}. We illustrate both experimental setups in Figure \ref{ABeffects}. 

In conventional quantum electrodynamics, the phase of matter couples to the potential $A_\mu$ through the gauge invariant Wilson loop $\Delta \Theta =  \oint q A_\mu \td x^\mu/\hbar$, where $x^\mu$ parametrizes a closed spacetime circuit. This coupling immediately implies that the gauge potential has physical significance even in regions of vanishing electric or magnetic fields. We can decompose the potentials into a gauge independent transverse magnetic component $\bs A_\text{T}$, and gauge covariant scalar electric components $A_0$ and $\bs \nabla\cdot \bs A$, that each mediate a distinct effect. The gauge invariant transverse mode of the vector potential induces a phase shift for space-like Wilson loops that encircle magnetic flux, 
\be\label{magnetic}
\Delta\Theta_{\text{Magnetic}}={e\over \hbar}\oint \bs A_\text{T}\cdot d\bs x={e\over \hbar} \Phi\,,
\ee
where $ \Phi=\int_{\Sigma} d\bs S\cdot \bs B$ denotes the magnetic flux through the surface $\Sigma$ enclosed by the Wilson loop. The gauge covariant scalar and longitudinal components induce a phase shift for time-like Wilson loops over time dependent potentials. In Coulomb gauge we have 
\be\label{electric}
\Delta\Theta_{\text{Electric}}=-{e\over \hbar}\oint A_0 \td t\,.
\ee
These unambiguous predictions have been made decades ago.

The phase shifts (\ref{magnetic}) and (\ref{electric}) depend only on the  gauge potential along the particle trajectory, and can be non-zero even when the particles never traverse any non-vanishing field strengths. Aharonov and Bohm therefore suggest to guide phase coherent beams of charged particles along two paths with different and non-trivial potentials, but vanishing fields. In order to verify the transverse mode of the potential, they propose to split an electron beam into two parts with each passing by a long solenoid on opposite sides. The interference pattern of the two beams will depend on the relative phase difference $\Delta \Theta_\text{Magnetic}$, which is proportional to the magnetic flux enclosed by the two beams. This magnetic Aharonov-Bohm effect has been experimentally observed soon after its prediction \cite{1960PhRvL...5....3C,1962NW.....49...81M}. The obvious electric counterpart of this experiment again employs two coherent electron beams, but now they are going on opposite sides of a capacitor (and through Faraday cages). The bias voltage is zero until the electron  wave-packet is well within the Faraday cages. The potential difference between the cages then ramps up for some finite time, but approaches zero again before the wave-packets exit the Faraday cages. Observing a relative phase shift $\Delta \Theta_\text{Electric}$ in the resulting interference pattern  would allow to verify the physical significance of the electric potential via (\ref{electric}), because the electrons never traversed non-vanishing electric fields. This electric Aharonov-Bohm effect has never been observed.

The coupling between the electric potential and the matter phase arises because of the interaction term in the action that couples the current to the vector potential, $S_\text{M}\supset -\int \td^4x A_\mu j^\mu$. The addition of the boundary term $\partial_\mu(A_\mu F^{\mu\nu})$ eliminates this coupling when Gauss' law is satisfied, which now contains only the transverse mode $S \supset \int \td^4x \bs A_\text{T}\cdot\bs j$, as in (\ref{electricd}). It is therefore intuitive that reduced quantum electrodynamic will yield only a magnetic, but no electric Aharonov-Bohm effect. We will confirm this intuitive expectation by explicitly solving the equations of motion in \S\ref{ABJ}. The  observation of the electric Aharonov-Bohm effect would thus rule out  reduced electrodynamics, while a null-observation would be inconsistent with conventional  QED.

\subsection{Non-relativistic charged scalar}\label{nrcs}
Rather than using electron beams, we propose in \cite{Bachlechner:2019deb} to use the superconducting order parameter in a test of the electric Aharonov-Bohm effect. 
The low-energy description of the superconducting order parameter is provided by the effective theory of a massive, complex non-relativistic scalar field $\phi$. Let us begin with the Lagrangian density of an uncharged field,
\be
{\cal L}_0=i\phi^*\dot\phi-{1\over 2m}\bs \nabla\phi^*\cdot \bs \nabla\phi-U(|\phi|)\,,
\ee
where $U(|\phi|)=\alpha|\phi|^2+\beta |\phi|^4/2$ is a potential and $m$ is the scalar field mass. Promoting the derivatives to be covariant under the spatial local $U(1)$ symmetry (\ref{spatialu1}) of reduced quantum electrodynamics, $\bs \nabla\rightarrow \bs D\equiv \bs \nabla-iq \bs A_\text{T}$, we find the conserved current
\be\label{scalarcurrent}
j^\mu=\left(q|\phi|^2,{i q\over 2m}(\phi\bs \nabla\phi^*-\phi^*\bs \nabla\phi)-{q^2 |\phi|^2\over m}\bs A_\text{T}\right)^\mu\,.
\ee
Combing this theory of a  scalar with the action for reduced electrodynamics in (\ref{Lred}), we find the Lagrangian for a charged non-relativistic scalar field
\be\label{scalarlagr}
L_{0}=\int_{\Sigma} \td^3x~~\bs A_\text{T}\cdot \dot{\bs E}_\text{T}+i\phi^*\dot\phi-\left[{1\over 2m}|(\bs \nabla-iq \bs A_\text{T})\phi|^2+U(|\phi|)+{\bs E^2(|\phi|^2)+\bs B^2(\bs A_\text{T})\over 2}\right]\,.
\ee
Of course, we could also have added the boundary term $d(A\wedge *F)$ to the Lagrangian density of conventional scalar electrodynamics and performed the Hamiltonian reduction as we did above for spinor electrodynamics in Appendix \ref{app1}, but this would merely constitute extra paperwork yielding the same result. 

The variation of the action with respect to the electric field and vector potential gives the usual Maxwell equations, while the variation with respect to the scalar gives a dynamical equation for charged matter
\be\label{scalareqm}
\int_{\Sigma} \td^3x~i\dot \phi=H_{\phi}\phi\,,
\ee
where we defined an effective Hamiltonian for $\phi$
\be\label{hamchargeds}
H_{\phi}\equiv \int_{\Sigma} \td^3x~{1\over 2m}(i \bs \nabla +q \bs A_\text{T})^2+q {\delta \over\delta  j^0}\left(U(|\phi|)+{\bs B^2+\bs E^2(j^0)\over 2}\right)\,.
\ee
Crucially, in this Hamiltonian the charges couple to electromagnetism through the  field strength $\bs E(j^0)$ and transverse potential $\bs A_\text{T}$, only, they do not couple to any scalar potential. As expected for a charged scalar in an external electric field, it is not possible to derive local equations of motion, since the electric field identically satisfies Gauss' law. 

The equations of motion (\ref{scalareqm}) are not the conventional equations of motion for a charged non-relativistic scalar. The difference stems from the fact that we imposed Gauss' law on the Lagrangian, before evaluating the equations of motion for the remaining physical degrees of freedom.\footnote{However, for vanishing boundary electric fields we can integrate by parts and recover the conventional dynamical equations. Only for non-vanishing boundary fields do the physical dynamics differ.} The electric field  contributes to the equations of motion for the phase of $\phi$ via the coupling
\be\label{voltage}
q {V}(x)\phi\equiv q \phi {\delta \over \delta j^0}\left({\bs E^2(j^0)\over 2}\right)\,,
\ee
where, following Weinberg\footnote{See section 21.6 of \cite{Weinberg:1995mt}.}, we defined the gauge invariant voltage ${V}(x)$ as the change in the energy density per change in charge density. In reduced electrodynamics matter couples to the voltage, not an electric potential. We can decompose the complex scalar field as $\phi\equiv \sqrt{n}e^{i\theta}$ to see that  the coupling (\ref{voltage}) has no imaginary part and therefore contributes only to the equations of motion for the phase $\theta$ of the complex scalar field. The dynamical equations of motion for all other physical fields coincide with those of conventional scalar quantum electrodynamics in Coulomb gauge.\footnote{While only some of the equations of motion differ, the resulting dynamics of all variables are affected because the equations are coupled.} In the following  subsections we will see that the modified dynamical equation for the phase is responsible for eliminating the  electric Aharonov-Bohm effect.

\subsection{Superconductors}\label{supconds}
Let us briefly discuss the effective low-energy description of superconductivity in reduced electrodynamics. 
We will loosely follow Weinberg's discussion of the conventional theory \cite{Weinberg:1995mt}. We have seen in \S\ref{sqed} that the action for quantum electrodynamics is invariant under a local spacetime $U(1)$ transformation that vanishes at the boundary, as well as an arbitrary local, but purely spatial $U(1)$ transformation,
\be
\{\psi,\,\bs A_\text{T}\}\rightarrow \{e^{-i e \lambda(\bs x)}\psi,\, \bs A_\text{T}+\bs\nabla\lambda(\bs x)\}\,.
\ee
This invariance guarantees that the Lagrangian takes the form (\ref{scalarlagr}),
\be
L_{0}=\int_{\Sigma} \td^3x~~\bs A_\text{T}\cdot \dot{\bs E}_\text{T}+i\phi^*\dot\phi-{\cal H}_{0}\left[\bs E_\text{T},\bs \nabla\times \bs A_\text{T},|\phi|, (\bs \nabla-iq \bs A_\text{T})\phi\right]\,.
\ee
We decompose $\phi=\sqrt{n}e^{i\theta}$ into a   phase $\theta$ and a density $n=j^0/q$, with $q=-2e$ the Cooper pair charge, such that after integrating out the density we have a theory for the phase
\be
L_{0}=\int_{\Sigma} \td^3x~~\bs A_\text{T}\cdot \dot{\bs E}_\text{T}-n \dot{\theta}-{\cal H}_{0}\left[\bs E_\text{T},\bs \nabla\times \bs A_\text{T},n,\bs \nabla \theta+q \bs A_\text{T} \right]\,.
\ee
Assuming that the system is in its stable ground state, this Lagrangian implies that the energy has a local minimum at $q \bs A_\text{T}=-\bs \nabla \theta$. $\bs A_\text{T}$ is divergence free, so the phase is coherent within the superconductor, $\theta(t,\bs x)=\theta(t)$, and magnetic fields are expelled from deep within a superconductor, $\bs B=\bs \nabla\times \bs A_\text{T}=0$.  With the current (\ref{scalarcurrent}) we arrive at what coincides with the conventional London equation in Coulomb gauge, 
\be
\bs j=-{nq^2\over m}\bs A_\text{T}\equiv-{\bs A_\text{T}\over \lambda_\text{L}^2}\,,
\ee
where we defined the London penetration depth $\lambda_\text{L}^2\equiv m/ nq^2$. From the Lagrangian we see that the negative number density $-n$ is the canonically conjugate momentum to the  phase $\theta$. The dynamical equation for the phase is
\be\label{secondj}
\dot{\theta}=-q{\delta {\cal H}_{0}\over \delta j^0}=-{q {V}\over \hbar}\,,
\ee
where we temporarily reinstated $\hbar$ and that we recognize as the second Josephson relation, but with a gauge invariant voltage $ {V}$ instead of the electric potential $A_0$, just as in Weinberg's discussion. The relation (\ref{secondj}) shows that for time independent fields superconductors carry current at vanishing voltage difference, and thus have no resistance.

While the phase is coherent within a superconductor, there can be a phase difference between two nearby, but disconnected superconductors. In the absence of electric and magnetic fields the  equation for $\phi$ in the ground state can be written as
\be\label{gleq}
-\zeta^2 \nabla^2\phi=\phi\,,
\ee
where we defined the material dependent coherence length $\zeta$. Consider now two superconductors of identical Cooper pair density $n$, but time dependent phase difference $\Delta \theta(t)$, separated by a distance $\delta$ at a junction of surface area $\bs S$. Solving (\ref{gleq}), and evaluating (\ref{scalarcurrent}) to find the total current between the superconductors $I=\bs j\cdot \bs S$ we have
\be\label{firstj}
I(t)=I_c \sin[\Delta \theta(t)]\,,~~ I_c={\hbar q n |\bs S | \over m \zeta \sinh(\delta/\zeta)}\,.
\ee
This is the first Josephson relation and describes a phase dependent DC current at zero voltage difference. It allows to deduce the relative phase (modulo $2\pi$) of two superconductors separated no more than the coherence length $\zeta$  by measuring the Josephson current.

\subsection{Testing the electric Aharonov-Bohm effect via Josephson currents}\label{ABJ}

\begin{figure}
  \centering
  \includegraphics[width=1\textwidth]{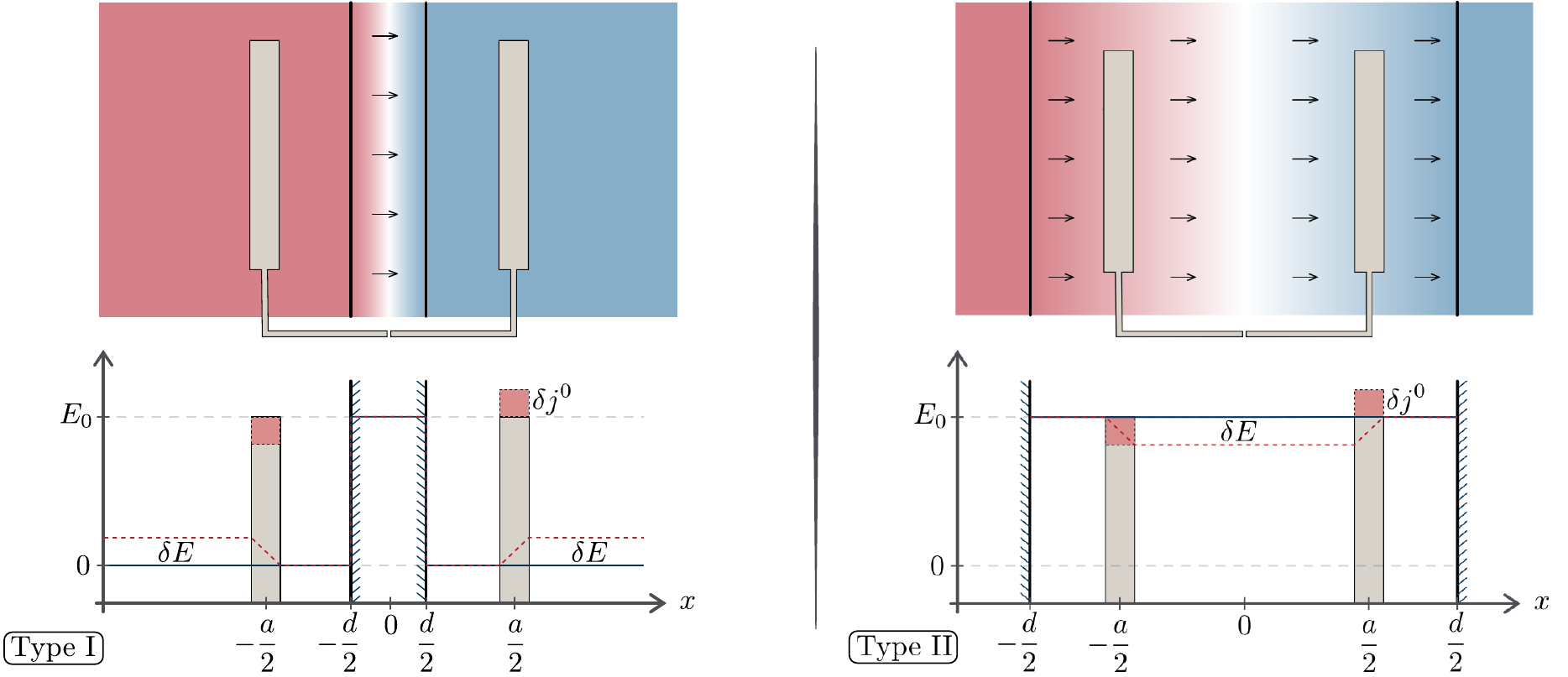}
  \caption{\small Top: Schematic illustration of superconductor interference experiment to detect the original (type I, left) and the field dependent (type II, right) electric Aharonov-Bohm effects. Bottom: Background electric field $E(x)$ (blue) and variation of the electric field $E(x)+\delta E$ (dashed red) with respect to the superconductor charge densities $j^0$, given fixed voltages between the capacitor plates.}\label{experiment}
\end{figure}

We have now reviewed all the necessary ingredients to understand a feasible physical setup \cite{Bachlechner:2019deb} that can conclusively test for the existence of the archetypal electric Aharonov-Bohm effect. The experimental setups are shown in Figure \ref{experiment}. Two superconducting nodes are separated by a distance $a$ and connected via thin superconducting wires that terminate at an insulating junction far away from the superconductors. The relative order parameter phase $\Delta \theta=\theta_2-\theta_1$ between two superconducting nodes is observable ($\text{mod}~2\pi$) via the  DC Josephson current $I$ in (\ref{firstj})   at zero voltage difference,
\be
\Delta \theta=\text{arcsin}\left[{I\over I_\text{c}}\right]\,,
\ee
where the critical  current $I_\text{c}$ depends on the details of  junction. In one setup a voltage difference between the superconductors is induced by placing planar capacitor plates separated by a distance $d>a$ on either side of the two superconductors (see right panel of Figure \ref{experiment}). This applies an electric field at the superconductors, shifts the relative energy of the Cooper pair condensates and allows to verify the already experimentally observed type II electric AB effect. In another setup the voltage difference is induced by placing the capacitor plates separated by $d<a$ in between the two superconductors and using further shielding to ensure a constant voltage at the superconductor (see left panel of Figure \ref{experiment}). In this setup the local electric fields vanish, and only the electric potential could interact with the superconductors, which allows to test for the first time the yet unobserved type I electric AB effect. In order to observe the phenomenon, the relative phase is initially recorded and subsequently a bias voltage $V=d\times  E_0$ is applied to the capacitor plates for some small time $T$. Comparing the relative phase via the DC Josephson effect before and after the application of the bias voltage yields the relation between the phase velocity and the electric potential of the superconductors, in both the type I and II experiments.

The prediction of conventional quantum electrodynamics for this experiment is unambiguous and standard \cite{Susskind:2015hpa,Bachlechner:2019deb}. The relative phase velocity is given by the conventional second Josephson relation, yielding a non-vanishing relative phase shift  of 
\be
\Delta \theta=-{q  T\over \hbar } \int_{-a/2}^{a/2} E(x)\, \td x\,,
\ee
for either experimental setup, where we used that the Coulomb potential difference between the superconductors is the integral of the electric field. There is no difference in the predictions between the type I and type II  AB effects.

While the prediction of reduced quantum electrodynamics for the type II experiment is identical to that of the conventional theory, we will now show that it predicts the absence of the type I electric AB phenomenon. This is intuitive, since the reduced theory does not contain an electric potential that could mediate a local interaction with the Cooper pairs.

Recall the dynamical equations governing the relative phase of the order-parameters $\phi_{1,2}$ of two superconducting nodes with non-trivial electric fields between them. In one spatial dimension, there are no transverse radiation modes $\bs E_\text{T}$ or $\bs A_\text{T}$, and the dynamical equation for a homogeneous, charged scalar field (\ref{scalareqm}) becomes
\be\label{sigmadot}
\int_{\Sigma} \td x~i\dot \phi= \int_{\Sigma} \td x~q {V}(x)\phi\,,
\ee
where we again have the gauge invariant voltage
\be\label{voltage2}
{V}(x)\equiv{1\over 2} {\delta \over \delta j^0}E^2(x)\,.
\ee
The non-local  equation (\ref{sigmadot}) allows to solve for the response of the phase velocity to all  variations of the charge density $j^0$ consistent with the boundary conditions. Within each superconducting node the Cooper pair charge densities $j^0_{1,2} = qn_{1,2}$ are constant, and vanish outside of the nodes. The voltage depends on the variation of the squared electric field $E^2(x)$ with respect to the density of the scalar fields $n_{1,2}$. The electric field is defined through Gauss' law,
\be
\partial_x E=j^0\,,\label{onedimgauss}
\ee
so it depends only on the charge density and the  value of the electric field at one point. The only distinguished points in our experimental setup are the capacitor plates, across which we can apply a fixed bias voltage $\Delta  V$, so in (\ref{onedimgauss}) we will fix the electric field acting on the capacitor plates to be  $E_0=\Delta V/d$, as illustrated in Figure \ref{experiment}.

Instead of considering the most general solution, we are interested only in the variations $\delta j^0$ of the charge density that generate the equations of motion for the relative phase. Increasing the Cooper pair density homogeneously by $\delta n$ at one node, and simultaneously decreasing it by the same amount at the other node gives the dynamical equation for the relative phase $\Delta \theta$,
\be
-\sqrt{n} {\cal V}_\text{node} \Delta \dot\theta=\int_\Sigma \td x ~{\delta\over \delta (n_2-n_1)} {E^2\over 2}\,,\label{relphasevel}
\ee
where $n$ is the on-shell Cooper pair density that corresponds to a charge  that is precisely canceled by the   background charge  density to give net neutral superconducting nodes. Using (\ref{onedimgauss}) and recalling that the experiment fixes the voltage (or electric field) acting on the capacitor plates, we find the variation of the   field strength under the relevant variation of the densities in Table \ref{fieldvariation}. Note that in the type I experiment the electric field varies with the charge density only in regions of vanishing background field, while in the type II experiment the electric field varies only in regions of non-vanishing background field. Since the electric field appears quadratically in the relative superconductor voltage, the relevant voltage vanishes in the type I experiment, while it coincides with the conventional electric potential difference in the type II experiment. Explicitly evaluating the relative phase velocity (\ref{relphasevel}) we find the total phase shift
\be
\Delta  \theta=\begin{cases}0~~~~~~~~~~~~~~~~~\,~~\text{for Type I experiment}\\ -{q  }a E_0 T/\hbar~~~~~~~\text{for Type II experiment}\end{cases}\,.
\ee
Under these boundary conditions, reduced quantum electrodynamics predicts the absence of the type I electric AB effect, while the type II effect is unaffected. These predictions are apparently consistent with observations \cite{mysteryreference,PhysRevB.40.3491,cite-key,PhysRevB.67.033307}.
\begin{table}\centering
\begin{tabular}{@{}p{3cm}cp{.5cm} p{3cm}c@{}}\toprule
\multicolumn{2}{c}{Type I experiment} && \multicolumn{2}{c}{Type II experiment} \\  \multicolumn{1}{c}{Location}&\multicolumn{1}{c}{~~Electric Field~~} &&\multicolumn{1}{c}{Location}&\multicolumn{1}{c}{~~Electric Field~~}\\ \cmidrule{1-2} \cmidrule{4-5}  
 \hspace{32pt}$x<-{a\over 2}$ & $q  {\cal V}_\text{node}\delta n$&&  \hspace{28pt} $x<-{d\over 2}$ & 0  \\
 $-{a\over 2}<x<-{d\over 2}$ &  0&  & $-{d\over 2}<x<-{a\over 2}$ &  $E_0$   \\
 $-{d\over 2}<x<{d\over2}$ &   $E_0$&& $-{a\over 2}<x<{a\over2}$ &  $E_0 -q  {\cal V}_\text{node}\delta n$ \\
  \hspace{9.3pt}${d\over 2}<x<{a\over2}$ &   $0$&&  \hspace{9.3pt}${a\over 2}<x<{d\over2}$ &   $E_0$   \\
    \hspace{9.3pt}${a\over 2}<x $ &   $q  {\cal V}_\text{node}\delta n$&&\hspace{9.3pt}${d\over 2}<x $ &   $0$   \\
\bottomrule
\end{tabular}
\caption{\label{fieldvariation}Variation of the electric field with the Cooper pair charge density mode that generates the dynamics of the phase velocity in the type I and type II setups, see Figure \ref{experiment}.}
\end{table}

Let us emphasize that the boundary conditions in the reduced theory crucially differ from those of conventional quantum electrodynamics: we do not fix the potential at ``infinity'', or any other undistinguished point. To fix the potential at infinity, we would require some ideal conductor to ensure vanishing field variations there. Instead, we fix the gauge invariant electric field across capacitor plates, as these boundary conditions may be easy to implement in a (finite) lab, for example by connecting a battery to the capacitor.  We do not claim that these boundary conditions are realized in nature, but merely that they are internally consistent and will be tested (and possibly ruled out) by observing the type I electric AB effect. Conversely, if this effect remains observationally elusive, this would provide useful information for a refined understanding of allowed boundary conditions in gauge theories.

\section{Semi-classical resonances}\label{semiclassicalr}
In constrained theories the on-shell action can be sensitive to total derivative (i.e. boundary) terms. For example, a term $\bs \nabla\cdot \bs E$ in the Lagrangian density appears to be a total derivative, but it evaluates to $j^0$ for classical solutions and clearly contributes to the action integral. Whenever the classical action of a constrained theory is relevant, it is therefore imperative to ensure that the action includes the correct boundary terms. The on-shell action for reduced QED differs from the action of QED. We now discuss how this affects the semi-classical approximation of the propagator at the simple example of electron pair production in external electric fields. This discussion has immediate relevance for semi-classical gravity.

The massive Schwinger model describes a massive Dirac particle of charge $e$ in $1+1$ dimensions \cite{Schwinger:1962tp,Coleman:1975pw,Coleman:1976uz}. For background fields exceeding $e/2$, the electric field becomes metastable and Schwinger pair creation allows for a (partial) discharge of the field strength. Tunneling and the decay of metastable states is a famously delicate subject in quantum field theories, and presents unresolved issues associated to the measure in  quantum gravity. While the semi-classical WKB technique that assigns tunneling probabilities between an initial ${\cal I}$ and final state ${\cal F}$ of
\be\label{wkbprob}
{\cal P}\propto e^{\frac{2i}{{\hbar}}\int_{\cal I}^{\cal F} \td S} \left[1+ \mathcal{O}(\hbar)\right]\,,
\ee
has been applied with great success to unconstrained theories, for gauge theories it is important to recognize the associated subtleties in order to arrive at the correct result. It is well known that  the exponent  of the semi-classical electron pair creation rate is not simply given by twice the Euclidean Maxwell action of the instanton solution, but that the non-vanishing boundary term $d(A\wedge * F)$ has to be included in the Lagrangian density \cite{Aurilia:1980xj,Brown:1987dd}.\footnote{A similar problem has been encountered for the nucleation process of charged black holes \cite{Brown:1988kg,Hawking:1995ap,Brown:1997dm}. Classical electrodynamics contains an electric-magnetic duality, so the nucleation rate of magnetic and electrically charged black holes is expected to be identical. However, the Maxwell action breaks this symmetry as it contains the term $(\bs E^2-\bs B^2)/2$, and one may fear runaway black hole production. Hawking and Ross argued that for the scenario of electric black hole production, again, the boundary term $d(A\wedge * F)$ should be included in the Maxwell action, while no such term is required for the magnetic process.} With this boundary term the on-shell action coincides with that of reduced electrodynamics. It is often argued that the inclusion of this boundary term is required for the consistency of the variational principle. This argument is at least partially unsatisfying, since the Maxwell action (when supplemented with gauge fixing) already yields a consistent variational principle for fixed potentials at the boundary. However, we have seen in \S\ref{FJacs} that the path integral measure becomes trivial when the canonical symplectic form of the theory contains  the  gauge invariant and unconstrained degrees of freedom. So rather than demanding just any consistent variational principle, we will see at the example of the Schwinger model that if the variational problem is well posed for boundary conditions for gauge invariant phase space variables, then the semi-classical propagator is well approximated by the exponential of the  action. Otherwise the non-trivial integration measure has to be accounted for.

The observations in this section explicitly illustrate that the Euclidean action does not always trivially relate to the semi-classical path integral. For quantum electrodynamics the Euclidean action alone is useful for semi-classical physics only when the variational principle is well posed for gauge invariant phase space variables. In the theory of Einstein gravity, the equivalent to the Maxwell action is the Gibbons-Hawking-York action. The fact that the exponential of the Maxwell action fails to describe semi-classical processes in quantum electrodynamics  might inform  whether its equivalent, the GHY action, could be valid to describe  quantum gravity.

\subsection{Resonances in semi-classical quantum mechanics}

\begin{figure}
  \centering
  \includegraphics[width=.6\textwidth]{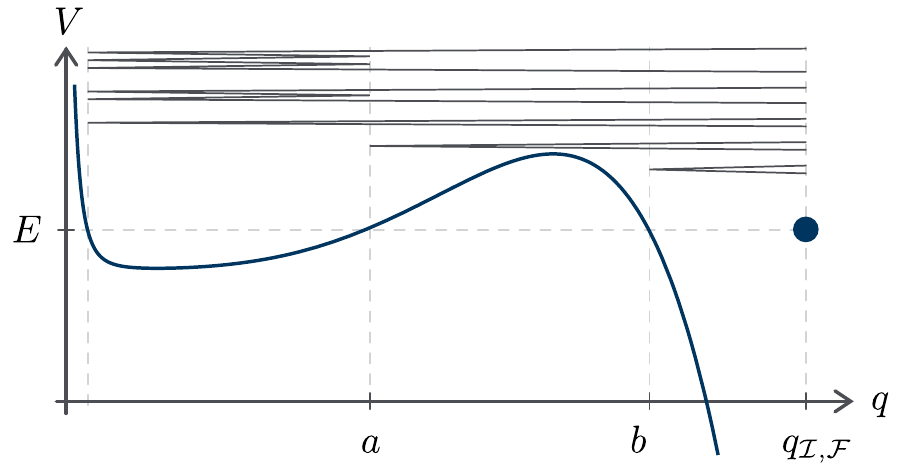}
  \caption{\small Example of a classically confining potential $V$ for a non-relativistic particle of energy $E$. The turning points are located at positions $q=a,b$, and some (complex) solutions to the classical equations of motion are indicated.}\label{potfigure}
\end{figure}
In this section we sketch the derivation of the tunneling probability (\ref{wkbprob}) for unconstrained quantum mechanical systems from the  semi-classical path integral.\footnote{The author thanks Kate Eckerle, Ruben Monten and Frederik Denef for many useful discussions associated with the unpublished notes on a related subject \cite{bemnotes}.} We closely follow the discussion in \cite{Holstein:1992da}.

The canonical phase space for an unconstrained mechanical system is parametrized by a variable $q(t)$ with conjugate momentum $p(t)$. The symplectic form is canonical, $\Omega=\delta p\wedge \delta q$, and the Lagrangian is given by
\be
L=p\dot{q}-H(p,q)\,.
\ee
To be explicit, let us assume the Hamiltonian of a non-relativistic particle of mass $m$ that is moving under the influence of a potential $U(q)$,
\be
H={p^2\over 2m}+U(q)\,.
\ee
The potential is illustrated in Figure \ref{potfigure} and has a metastable minimum at $q<a$ and is unbound at large $q>b$, where $a$ and $b$ denote classical turning points. We are interested in the rate at which particles of energy $E$ tunnel out of the metastable minimum. This problem is  the quantum mechanical analog of vacuum decay (or flux discharge) in quantum field theory. Clearly, the rate at which particles escape though the potential barrier of height $V_\text{max}$ depends on the initial state: particles of energy $E>V_\text{max}$ will escape unimpeded, while particles of energy $E< V_\text{max} $ will have to undergo quantum tunneling. In order to avoid an initial vacuum state where the WKB approximation breaks down we will consider a scattering process in which an unbound in-going particle at $q_{\cal I}=q(t=0)>b$ with energy $E<V_\text{max}$ scatters off the potential barrier and returns to $q_{\cal F}=q(t=T)>b$. The initial state, as well as possible trajectories between the initial and final states are illustrated in Figure \ref{potfigure}. The late-time correlation function
\be
U_{T}(q_{\cal I}, q_{\cal F};0,T) \equiv \braket{q_{\cal F},T |  q_{\cal I},0}\,,\label{correlator}
\ee
will be dominated by the decay of the metastable state that is populated in the scattering process. Our task is therefore to evaluate the correlation function (\ref{correlator}) at late times.

Given that the Hamiltonian $H$ generates time translations, we have the path integral for the causal propagator 
\be
\label{propagator}
U_{T}(q_{\cal I}, q_{\cal F};0,T) = \braket{q_{\cal F} |\theta(T) e^{-\frac{i}{\hbar} H T} | q_{\cal I}} = \theta(T)\int_{q_{\cal I}}^{q_{\cal F}}{{{\cal D} q} \, {\cal D} p~ \exp\left[{\frac{i}{\hbar} \int_0^{T}{  p \dot{q} - H(p, q)~\td t}}\right] }\,,
\ee
where the path integral is performed over all paths connecting initial and final states. Since the momentum appears quadratically in the Hamiltonian we can evaluate the integral over $p$, and find in the limit $\hbar\rightarrow 0$  the semi-classical propagator
\be
U_{T}(q_{\cal I}, q_{\cal F};0,T) \propto \exp\left[-i \left(\int^{q_{\cal F}}_{q_{\cal I}} \td q ~p- H_\text{cl} T\right)\right]\,,
\ee
where the exponent contains the action evaluated on the classical trajectory, $\delta S=0$, and $H_\text{cl}$ is the  Hamiltonian on that trajectory. In order to find the resonances corresponding to metastable states, we are  interested in the fixed energy Fourier transform of the propagator,
\be
U(q_{\cal I}, q_{\cal F};\omega+i\epsilon) \equiv \int \td T\, U_{T}(q,0; q,T)e^{i(\omega+i\epsilon)T}=\bra{q_{\cal F}} {i\over \omega+i\epsilon-H}\ket{q_{\cal I}}\,,\label{semiclassicalT}
\ee
and we introduced the small positive parameter $\epsilon$ to ensure convergence of the integral. Performing the integral over time yields the frequency space propagator
\be\label{series}
\bra{q_{\cal F}} {i\over \omega+i\epsilon-H}\ket{q_{\cal I}}={\cal C } \times \sum_{ I = \left\{\substack{H_\text{cl}=\hbar \omega + i\epsilon \\ (q_{\cal F},q_{\cal I})} \right\}} \mu_I \, \exp\left[{\frac{i}{\hbar} \int_{q_i}^{q_f} \td q\,p}\right]\,,
\ee
where the path integral is performed over classical paths of fixed Hamiltonian $H_\text{cl}$,  labeled by $I$. ${\cal C }$ denotes an irrelevant prefactor and $\mu_I$ are contributions from turning points where the semi-classical approximation fails. Figure \ref{potfigure} illustrates some of the paths contained in the infinite series (\ref{series}). Performing the sum over all semi-classical paths yields a series of poles in the propagator corresponding to all (metastable) bound states. Expanding around a pole at energy $H_\text{cl}$ gives the frequency domain Breit-Wigner distribution
\be
|U(q_{\cal I}, q_{\cal F};\omega+i\epsilon)|^2\propto {1\over (\omega-H_\text{cl})^2+ \Gamma^2/4}\,,
\ee
which corresponds to the time domain Fourier transform $\propto e^{-\Gamma t}$ indicating a decay process with decay rate
\be
\Gamma={\cal A} \exp\left[{{2 i\over \hbar}}\int_{a}^b \td q~ p\big |_{H=H_{\text{cl}}+i\epsilon} \right]\,.\label{crate}
\ee
The integral is performed over the outgoing ($\text{Re}[\dot{q}]>0$) classical path of least absolute action between the turning points with slightly positive imaginary Hamiltonian. 

The classical momentum for our mechanical model above is given by 
\be
p_\text{cl}=\eta \sqrt{2m [H_\text{cl}-U(q)]}\,,
\ee
where $\sqrt{\dots}$ denotes the positive branch of the square root and $\eta=\pm 1$. Using Hamilton's equations we find the requirement for an outgoing solution at the classical turning point
\be
\text{Re}[\dot{q}]\big|_{q=a,b}=\text{Re}\left[{\partial p_\text{cl}\over \partial H_\text{cl}}\right]^{-1}\bigg|_{q=a,b}\propto \eta \, \text{Re}\left[\sqrt{i\epsilon}\right]\propto\eta \,\text{Re}[1+i]>0\,,
\ee
such that continuity of the momentum with the outgoing solution in the classically allowed region selects the positive imaginary momentum solution within the  forbidden region, $\eta=+1$, and hence the familiar result without  any sign ambiguity in the exponent
\be
\Gamma={\cal A} \exp\left[{-{2\over \hbar}}\int_{a}^b \td q~ \sqrt{2m (V-H_\text{cl})}\right]\,.
\ee

Although we suppressed many details, we did cover some crucial aspects of the semi-classical derivation of decay rates that are not usually considered: we found that the semi-classical decay rate contains the path integral over the least action path parametrized by  unconstrained degrees of freedom between initial and final configurations at fixed and slightly positive imaginary classical Hamiltonian. It is important that the Hamiltonian coincides with the energy of the system. This last requirement fails for the Maxwell action evaluated in the Schwinger model, which invalidates its naive use in the semi-classical propagator.

\subsection{Semi-classical Schwinger pair creation in reduced QED}
Having reviewed the derivation of the  decay rate  from the semi-classical  path integral, we now turn to evaluate the  rate of electron pair production in an electric field by finding the action that allows us to evaluate the path integral (\ref{crate}).

\begin{figure}
  \centering
  \includegraphics[width=.6\textwidth]{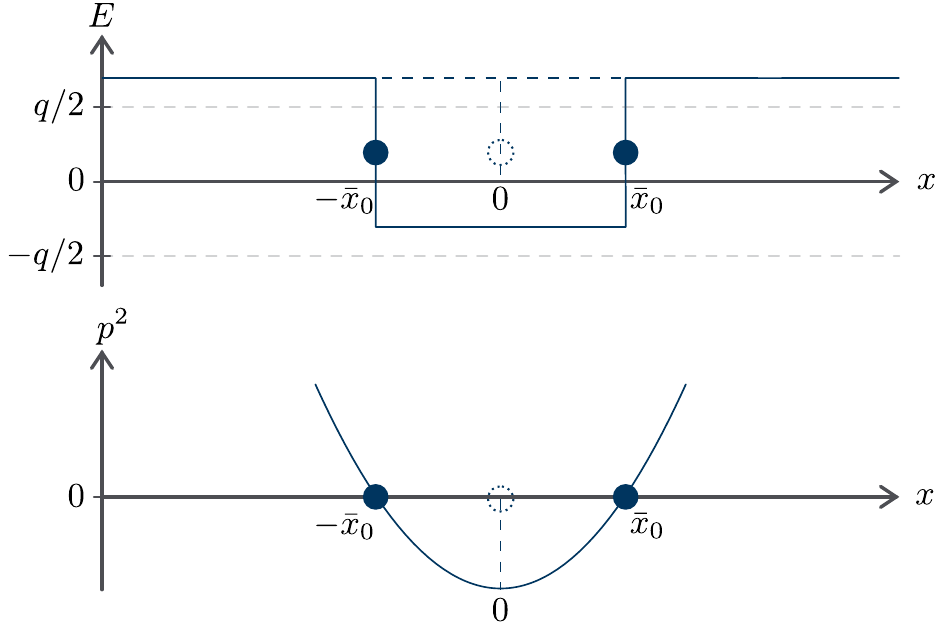}
  \caption{\small Schematic illustration Schwinger pair production. Top: Electric field as a function of position. Bottom: momentum squared as a function of position. The particles are shown by blue dots at the location of the turning point.}\label{schwinger}
\end{figure}

Consider a pair of particles with charge $\pm e$ in a constant electric field $E_0$. Explicitly, we consider a 1+1 dimensional setup on the spacetime manifold ${\cal M}$ with coordinates
\be
{\cal M}=\{\,(t,x)~:~t_{\text{I}}\le t\le t_{\text{F}}\,,~
-x_{\text{max}}\le x\le x_{\text{max}}\,\}\,.
\ee
The density of a particle of charge $e$ at $\bar{x}(t)$ is given by
\be\label{current}
j^0=e \delta[x-\bar{x}(t)]\,.
\ee
Gauss' law in one spatial dimension, $E'=j^0$, implies that the electric field is constant and changes by $\pm e$ at the particle locations. In order to study pair-creation we choose  boundary conditions that fix the electric field $E_0$ at any point $|x| >\bar{x}$, i.e.
\be
E(t,0)=E_0-e\,,~~E(t,x_\text{max})=E_0\,.
\ee 
These boundary conditions correspond to particles that evolve within a capacitor of fixed voltage. During pair production the particles travel along a classically forbidden trajectory from being coincident, $\pm\bar{x}(t_{\text{I}})=\pm 0$, to a classical turning point at $\pm\bar{x}(t_{\text{F}})=\pm \bar{x}_0$ from where the particles begin to evolve classically. We illustrate the nucleation process in Figure \ref{schwinger}.

Let us now evaluate the action for this process. Combining the Lagrangian for a relativistic point particle of mass $m$ with the Lagrangian (\ref{Lred}) of reduced electrodynamics we have the action for one particle
\be
S=\int  \left[p \dot{\bar{x}}-\sqrt{m^2+p^2}-\int_0^{x_\text{max}} {E(\bar{x})^2\over 2} \td x\right]~\td t\,.
\ee
Note that there are no transverse modes in one spatial dimension. In the reduced system the scalar mode of the electric field is defined via the solution of Gauss' law with the given boundary conditions,
\be
E(\bar{x})\equiv E_0-e \theta(|\bar{x}|-|x|)\,.
\ee
We can evaluate the spatial integral appearing in the action to find
\be
S=\int \td t ~p \dot{\bar{x}}-\big(\underbrace{\sqrt{m^2+p^2}-e(E_0-e/2)\bar{x}}_H\big)\,.
\ee
We dropped a constant and irrelevant term $+E_0^2 x_\text{max}/2$ and recognize the Hamiltonian $H$ for a single particle. Solving $H=0$ we find the classical momentum $p_\text{cl}$ and a turning point $\bar{x}_0$,
\be
p|_{\text{cl}}=\pm m\sqrt{\bar{x}^2/\bar{x}_0^2-1}\,,~~\bar{x}_0={m\over e(E_0-e/2)}\,.
\ee
We find the exponent appearing in the electron pair production rate from (\ref{crate}) by integrating $p_\text{cl}$ over the outgoing, classically forbidden trajectory. Including a factor of two for the two particles being nucleated this gives the Schwinger pair production rate
\be
\Gamma\propto \exp\left(2\times 2\times i\int_{0}^{\bar{x}_0}\td x~p_{\text{cl}}\right)=\exp\left[-{m^2\pi\over e(E_0-e/2)}\right]\,.
\ee
This is indeed the correct result \cite{Schwinger:1951nm,Kleban2011}.

\subsection{Comparison to the Maxwell instanton action}
In the previous subsection we have seen that the exponential of the reduced quantum electrodynamics action immediately relates to the physical Schwinger pair production rate. On the surface  $\td A= F$ this action differs from the Maxwell action by a term that contains only total derivatives,
\be
S -S_{\text{M}}=\int_{\cal M} \td(A\wedge *\td A) =\int_{\cal M} \td^2x\partial_\mu(A_{\nu}F^{\mu\nu})\,.
\ee
We now evaluate this term explicitly,
\bea
S -S_\text{M}&=&\int_{t_{\text{I}}}^{t_{\text{F}}}\td t \int_0^{x_\text{max}}\td x\left(\partial_x(A_t E)-\partial_t(A_x E)\right)\\
&=&\int_{t_{\text{I}}}^{t_{\text{F}}}\td t\left[A_t E\right]_0^{x_\text{max}}-\int_0^{x_\text{max}}\td x \left[A_x E\right]_{t_{\text{I}}}^{t_{\text{F}}}\,.
\eea
This boundary action does not only depend on fixed boundary data, as it contains both $A_\mu$ and derivatives of $A_\mu$. We can evaluate the boundary term in the Coulomb gauge $A_x=0$ to define the electric potential
\be
A_t(t,x)=-\int_0^x \td x'~E(t,x')\,.
\ee
The boundary terms are sensitive to only the boundary electric potential, but this potential is non-local and depends on the electric field configuration within the bulk of ${\cal M}$. The boundary action is therefore sensitive to the electric field within ${\cal M}$, not just  the boundary data. Explicitly we find 
\bea
S -S_\text{M}|_{A_x=0}&=&\int_{t_{\text{I}}}^{t_{\text{F}}}\td t\left[\left(-\int_0^x \td x'~E(t,x')\right) E(t,x)\right]_0^{x_\text{max}}\\
&=&\int_{t_{\text{I}}}^{t_{\text{F}}} \left[2\times e(E_0-e/2)\bar{x}-E_0^2 x_\text{max}\right]\td t\,,
\eea
where again the constant last term can be dropped but the first term is non-trivial. Using the Maxwell action alone in a semi-classical approximation to a path integral would therefore lead to a gauge dependent and incorrect result.

\subsection{Application to semi-classical quantum gravity}
The Maxwell action for electrodynamics is similar to the Gibbons-Hawking-York (GHY) action for gravity \cite{York:1972sj,Gibbons:1976ue}. While in Maxwell's action the potentials are fixed at the boundary, in the GHY action the boundary metric is fixed. Both approaches fix some of the gauge degrees of freedom, and the symplectic forms of both theories are not gauge invariant. We saw above that the exponential of the Maxwell action does not immediately relate to semi-classical propagators. It is therefore not intuitive that the GHY action does any better for gravity.

In contrast, the theory of reduced quantum electrodynamics has gauge invariant phase space coordinates and a canonical symplectic form, which renders the path integral measure trivial, and thus allows us to easily evaluate the correct semi-classical path integral. It is well known that the semi-classical path integral for quantum gravity using the GHY action yields many puzzles, including an ambiguous path integral measure. This observation suggests that that a reduced theory of gravity with gauge invariant phase space, canonical symplectic form and trivial path integral measure might yield a better behaved theory of semi-classical quantum gravity. The reduced gravitational action was constructed in \cite{Bachlechner:2018pqk} for a simple example of Jackiw-Teitelboim gravity \cite{Teitelboim:1983ux,Jackiw:1984je} that describes spherically symmetric four-dimensional systems \cite{Kuchar:1994zk}. Just as in Maxwell theory, the use of gauge invariant variables changes the instanton action relevant for the decay of metastable (de Sitter) states with non-vanishing asymptotic fields \cite{tbvacuumdecay} .

\section{Conclusions}\label{conclusions}

Faddeev and Jackiw proposed a reduced theory of  quantum electrodynamics that eliminates the scalar potential, is manifestly gauge invariant and consistent with observations \cite{Faddeev:1988qp,Jackiw:1993in}. The widely known magnetic Aharonov-Bohm (AB) effect, that serves to verify the significance of the transverse vector potential, was predicted and observed six decades ago \cite{Ehrenberg_1949,PhysRev.115.485,PhysRev.123.1511,1960PhRvL...5....3C,1962NW.....49...81M}. AB also predicted a less known electric phenomenon that would verify the physical significance of the scalar potential, but this effect has never been observed. The existence of this electric effect is a key assumption in recent theoretical investigations of the allowed boundary conditions in gauge theories \cite{Dirac:1955uv,Frohlich:1979uu,Balachandran:2013wsa,Strominger:2013lka,Balachandran:2014hra,Kapec:2015ena,Donnelly:2015hta,Hawking:2016msc,Strominger:2017zoo,Henneaux:2018gfi,Gomes:2018dxs,Gomes:2019xhu,Gomes:2019rgg,Gomes:2019xto,Harlow:2019yfa,Giddings:2019ofz}.  In the present work we showed that the unobserved electric AB effect is absent in the reduced theory of quantum electrodynamics. A conclusive experimental test is feasible via a simple superconductor interference experiment \cite{Bachlechner:2019deb}. A reduced theory of  gravity would have an impact on the predictions of semi-classical gravity \cite{tbvacuumdecay}, and might alleviate a problem with the measure. It will be interesting to understand the impact of Hamiltonian reduction on the status of the strong CP problem.

\section{Acknowledgements}
We thank  Robert Dynes, Dan Green, Raphael Flauger, Marc Henneaux, Matthew Kleban, Per Kraus, Liam McAllister and John McGreevy for useful discussions.  We particularly thank Kate Eckerle, Frederick Denef and Ruben Monten for discussions and collaboration on the unpublished notes \cite{bemnotes}, that in part inspired the underlying idea of the present work. This work was supported in part by DOE under grants no. DE-SC0009919 and by the Simons Foundation SFARI 560536.  This work was performed in part at the Aspen Center for Physics, which is supported by National Science Foundation grant PHY-1607611.

\appendix
\section{Hamiltonian reduction for spinor electrodynamics}\label{app1}
To illustrate the power of the Hamiltonian reduction, we now perform the procedure explicitly to re-derive (\ref{electricd}).

Just as above, we perform a Hodge decomposition on the compact manifold $\Sigma$ of the spatial components $\bs A$ and $\bs E$,
\be
\bs A\equiv \bs A_\text{T}+\bs \nabla A_\text{s}\,,~~\bs E\equiv \bs E_\text{T}+\bs \nabla E_\text{s}\,.
\ee
Substituting this into (\ref{electricaction}), dropping irrelevant boundary terms that do not affect the symplectic form and evaluating the non-linear constraint for the magnetic field, $0=\delta H/\delta \bs B$, we arrive at the the Lagrangian
\be
{ L}_{ \frac{1}{2}}=\int_{\Sigma} \td^3x~{1\over 2}\bs \zeta^\top\bs \omega\dot{\bs \zeta}-H(\bs \zeta, A_\text{s},E_\text{s},A_0)\,,
\ee
where
\be
\zeta=\left(\begin{array}{c}\bs A\\i\psi^\dagger\\\bs E\\\psi\end{array}\right)\,,~~\bs \omega=\left(\begin{array}{cc}0&\bs 1\\- \bs 1&0 \end{array}\right)\,,
\ee
and the Hamiltonian is now given by
\be
H=\int_{\Sigma}\td^3x{\bs E^2+\bs B^2(\bs A_\text{T})\over 2}-\bar{\psi}\left(\bs \gamma\cdot \left[i\bs \nabla+ e \bs A_\text{T}+e\bs \nabla A_\text{s}\right]-m\right)\psi+A_0(\psi^\dagger\psi-\bs \nabla\cdot {\bs E})\,.
\ee
Just as before, $A_0$ appears linearly in the Hamiltonian. The variation of this Lagrange multiplier therefore  imposes Gauss' law, the only true constraint of the theory. Using (\ref{essol}) to solve for $E_\text{s}(j^0)$ thus removes the longitudinal mode of the electric field, $\bs E(\psi^\dagger\psi)\equiv \bs E_\text{T}+\bs\nabla E_\text{s}(\psi^\dagger\psi)$. However, note that the longitudinal mode of $\bs A$ appears in the symplectic form, so we have a non-canonical kinetic term in the Lagrangian,
\bea
{ L}_{\frac{1}{2}}&=&\int_{\Sigma} \td^3x~\bs A_\text{T}\cdot \dot{\bs E}_\text{T}+ \dot{A}_\text{s}\bs\nabla \cdot \bs E(\psi^\dagger\psi)+i{\psi}^\dagger\dot{\psi}\\&&~~~~~~~~~~-{\bs E^2(\psi^\dagger\psi)+\bs B^2(\bs A_\text{T})\over 2}+\bar{\psi}\left(\bs \gamma\cdot \left[i\bs \nabla+ e \bs A_\text{T}+e\bs \nabla A_\text{s}\right]-m\right)\psi\,,\nonumber
\eea
where we integrated by parts the  kinetic term, and added irrelevant time derivatives. Using the solution for the longitudinal mode of the electric field, $\bs\nabla\cdot \bs E(\psi^\dagger\psi)=e\psi^\dagger \psi$, and performing the Darboux transformation $\psi\rightarrow e^{ie A_\text{s}}\psi$ then results in a canonical kinetic term and the longitudinal mode $A_\text{s}$ entirely disappears from the theory, yielding just the action (\ref{electricd}) that we already argued for earlier.  This derivation was first performed by Faddeev and Jackiw for vanishing boundary fields  \cite{Faddeev:1988qp,Jackiw:1993in}.

\bibliographystyle{JHEP}
\newpage
\bibliography{bubblerefs}
\end{document}